\documentclass[twocol]{ametsocV6.1}

\usepackage{caption}
\usepackage{subcaption}
\usepackage{float}

\usepackage{bm}
\ifdraft
\else
\PassOptionsToPackage{hyphens}{url}
\usepackage{hyperref}
\hypersetup{
	breaklinks,
	colorlinks=true,
	linkcolor=blue,
    urlcolor=purple,
	citecolor=purple,
 }
\fi

\title{Rapid topographic scatter of near-inertial waves generated by storms}

\authors{Ashley~J.~Barnes,\aff{a,b}\correspondingauthor{A.~J.~Barnes, ashley.barnes@anu.edu.au}
Callum~J.~Shakespeare\aff{a,b}, \ifdraft\\\else\fi
Andrew~McC.~Hogg,\aff{a,b,c} and 
Navid~C.~Constantinou\aff{c,d}}

\affiliation{
\aff{a}{Research School of Earth Sciences, Australian National University, Canberra, ACT, Australia}\\
\aff{b}{Australian Research Council Center of Excellence for Climate Extremes, Australia}\\
\aff{c}{Australian Research Council Center of Excellence for the Weather of the 21st Century, Australia}\\
\aff{d}{School of Geography, Earth and Atmospheric Sciences, University of Melbourne, Parkville, VIC, Australia}
}

\abstract{Internal waves propagate on the ocean's stratification, carrying energy and redistributing momentum through the ocean. When internal waves break, they contribute to diapycnal mixing in the ocean interior, but this breaking behaviour depends upon the scale of the waves. Low-mode internal waves have larger horizontal and vertical scales, and thus break less readily than higher-mode waves. The scattering of internal waves by topography is an important mechanism in transferring internal wave energy to smaller scales that are more conducive to wave breaking and mixing processes. In this study, we propose and investigate a mechanism in which storm-generated low-mode internal waves scatter at topography. We hypothesise that horizontally propagating internal wave modes generated by strong winds (i.e., due to a storm) can rapidly dephase; these dephased waves can then be scattered from topography, resulting in higher-mode upward-propagating waves within hours of the passage of a storm. We investigate this phenomenon in an idealised numerical model of a storm passing over a prominent ridge. Bottom-scattered near-inertial internal waves propagate away from the ridge rapidly in the wake of the storm. We perform several perturbation experiments varying the properties of the ocean, the winds and the topography. The bottom-scattered waves exhibit spatial downscaling, and have an energy flux equivalent to $10\%$ the magnitude of the energy flux from surface-generated near-inertial waves in our domain. Although small in a globally averaged sense, we argue that the topographic scatter of storm-generated near-inertial waves could account for the unexplained near-inertial wave signals found in ocean observations and numerical studies.}

\begin{document}

\maketitle

\statement{
The ocean is stratified: denser waters are in the deep ocean and lighter waters closer to the surface. 
How and where waters of different densities mix is important as this contributes to global ocean circulation and how the ocean transports heat from the tropics to the poles.
This ocean mixing is influenced greatly by the winds and tides that churn up the ocean near boundaries and generate internal waves, the breaking of which is thought to be the major source of turbulent mixing in the ocean interior. 
These waves propagate in the interior of the ocean rather than just at the surface, and are made possible by the ocean's stratification.
These waves are very difficult to observe and so our understanding of their generation, propagation and breaking is an active field of research. 
Here, we demonstrate that internal waves generated by strong storms, like typhoons or cyclones, can rapidly scatter from nearby topography at the sea floor.
Previous studies looked at waves generated by larger-scale winds rather than individual storms and suggested that the waves take weeks to reach the seafloor.
Here, we show that storm-generated waves scatter at the seafloor within hours.
Our study could help explain some of the bottom-intensified wave activity that has been observed within hours of the passage of a storm. 
We find that the waves propagating upwards from topography have smaller spatial scales, and up to $10\%$ the energy flux compared to the surface-generated waves that they scatter from. 
We argue that this rapid wave-scattering mechanism is worthy of further study to better understand how they might fit into the ocean mixing picture.}

\section{Introduction}
\label{section:introduction}

Internal waves, which manifest as mixed horizontal and vertical oscillations in a rotating stratified fluid, permeate the world's oceans. 
Internal waves are important to the fields of physical oceanography and climate science more broadly due to their significant contribution to ocean mixing and momentum transfer \citep{bell_topographically_1975,alford_near-inertial_2016,sarkar_topographic_2017,thomas_chapter_2022,musgrave_chapter_2022}. 
Global estimates of the energy contained by the internal wave field are supported by sparse \emph{in-situ} observations and satellite altimetry \citep{zhao_global_2016,whalen_large-scale_2018}. 
However, capturing the spatial variability of internal waves, and in turn their contribution to local episodic mixing, remains an ongoing challenge in oceanography, particularly in the deep ocean where observations are especially sparse. 
A further complication is that some types of internal wave can contribute to the local mixing at the generation site, while others can carry their energy great distances to enhance mixing in remote locations \citep{waterhouse_global_2014}. 

Ocean models can provide insights where observations are lacking, and accurately capturing internal wave induced mixing in numerical models is a large area of study. 
Some approaches to this problem, such as \cite{laurent_role_2002} and \cite{lavergne_parameterization_2020}, construct spatially varying maps of the wave field by parameterising different wave generation mechanisms. 
This method has shown promise in matching observations where they are available \citep{lavergne_parameterization_2020}, but an unavoidable source of error in such approaches comes from processes that are omitted. 
By estimating the importance of this missing physics, we can increase our understanding of how well wave-induced mixing is represented in numerical models.

The bulk of internal waves begin their life-cycle at the boundary of the ocean -- either at the surface, generated via winds \citep{thomas_chapter_2022}, or near the bottom as a result of flow over topography \citep{musgrave_chapter_2022}. 
In the first case, inertial oscillations at the surface are readily excited by the wind, which in turn generate near-inertial waves that can propagate downwards, beyond the mixed layer \citep{dasaro_energy_1985}.  These wind-driven, surface-generated near-inertial waves (NIWs) are responsible for between 0.3-1.1$\;$TW  of energy flux at the surface \citep{jiang_estimating_2005,rimac_influence_2013}. The exact mechanism by which the near-inertial energy escapes the mixed layer depends significantly on the length scale of the wind forcing. Forcing by small, intense storm systems \citep{gill_behavior_1984} generates low-modes which propagate thousands of kilometers horizontally away from their generation site \citep{alford_redistribution_2003}.  For larger scale winds, the near-inertial oscillations instead need to be downscaled by $\beta$-refraction or mesoscale interactions \citep{thomas_chapter_2022} in order to escape the mixed layer.  These waves propagate at very low vertical group speeds and thus shallow angles, with observational studies suggesting that vertical propagation takes weeks to bring such waves to the seafloor after generation \citep{kunze_observations_1984,ma_bottom-reached_2022}. 
Here, they may scatter, dissipate or reflect depending on the local topography \citep{igeta_scattering_2009}.

In the second case, topography generates internal waves directly. 
When deep ocean flows interact with topography, this can lead to vertical motions, and thereby the radiation of internal waves due to the restoring effect of the stratification. 
Examples of such interactions include the barotropic tides generating internal tides \citep{bell_topographically_1975}, as well as lee waves excited by geostrophic currents \citep{nikurashin_radiation_2010}, which are thought to have globally-integrated energy fluxes of 0.9-1.5$\;$TW and 0.05-0.85$\;$TW respectively \citep{nikurashin_global_2011,egbert_significant_2000,waterhouse_global_2014,bennetts_closing_2024}.

There is an indirect connection between wind-generated and topographically-generated internal waves. 
Wind forcing energises both long-lived and transient geostrophic currents \citep{wunsch_work_1998}, which in turn shed energy at the ocean bottom via topographic lee waves \citep{nikurashin_radiation_2010}. 
\cite{nikurashin_global_2011} estimated that $20\%$ of energy input to  geostrophic flows through synoptic scale wind forcing was converted to lee wave generation. 
Resonant interactions of the lee waves and geostrophic flows \citep{nikurashin_radiation_2010}  drive near-inertial energy and enhanced mixing near topographic features \citep{hu_dynamic_2020,brearley_eddy-induced_2013,liang_eddy-modulated_2012}. 
Given that the wind-driven geostrophic currents responsible for lee wave generation vary on timescales of months and years, this mechanism represents a persistent, steady source of internal wave generation.

In contrast, recent observations have found rapid generation of near-inertial waves at the seafloor within hours of a storm passing overhead. \cite{van_haren_challenger_2020} found evidence for enhanced mixing in the depths of the Mariana Trench, with a mooring detecting patterns in the stratification similar to that of breaking internal waves within hours of a category 4 typhoon. 
The rapid timescale and close proximity of the mixing to the storm suggest that neither of the two cases discussed above --- that is, the interaction of lee waves with geostrophic flows nor the downwards propagation of near-inertial waves with low vertical group speeds excited by large scale winds --- are able to account for the observations. 
As such, a new mechanism is needed.

In this work, we investigate the rapid scattering of near-inertial internal waves at the seafloor in the wake of a storm system.
We hypothesise that the sharp horizontal gradients of storm systems allow for the generation of full-depth vertical modes, unlike other near-inertial waves generated by larger scale wind patterns as discussed in \cite{kunze_observations_1984} and \cite{ma_bottom-reached_2022}.
Other full-depth internal wave modes, like the baroclinic tide, are known to scatter at topography, which can redistribute their energy to smaller horizontal scales \citep{peter_muller_scattering_1992,laurent_role_2002,buhler_decay_2011, kelly_coupled_2013}.
Like these other full-depth internal waves, we demonstrate that the near-inertial waves generated by a storm can also scatter from local topography, with upwards propagating internal waves appearing almost immediately.
This short timescale of the bottom response is of particular interest, as it makes this mechanism a contender for explaining the aforementioned enhanced bottom mixing coinciding with the typhoon. 

The mechanism for this rapid topographic scattering is the dephasing of vertical modes. \cite{gill_behavior_1984} studied the ocean response to a short-lived wind event, like the passage of a storm, that instantaneously accelerates the mixed layer and imprints onto multiple internal wave vertical modes.
The dephasing of these vertical modes results in the rapid communication of the surface layer perturbation downwards below the thermocline. 
Given the full depth nature of the vertical modes generated by storm systems, we extend Gill's work to consider a rapid response at the sea floor.

In the presence of topography, we anticipate that the deep oscillatory motions driven by dephasing vertical modes will generate internal waves in a similar fashion to internal tides \citep{bell_topographically_1975}. 
Like internal tides, this generation process can equally be thought of as the scattering of a large scale wave (barotropic tide, or surface-generated NIW) to a smaller scale wave (baroclinic tide, or topographically scattered NIW). 

As discussed above, the scattering of near-inertial waves generated by large-scale winds is associated with a significant time lag due to them not exciting full-depth modes, and the low propagation angle of the waves as they escape the mixed layer.
In contrast, the smaller horizontal scales of  storm-generated near-inertial waves, which are the focus of this study, readily excite full-depth modes, and in turn we expect a rapid bottom response as these modes dephase.
In the presence of sufficiently narrow and prominent topographic features, we further expect the waves to scatter to higher modes, facilitating energy downscaling which may ultimately lead to additional bottom mixing.

Given that storms convert large amounts of wind energy into surface layer oscillations and downward propagating near-inertial waves \citep{sanford_upper-ocean_2011,whalen_large-scale_2018}, the coincidence of a storm directly above topographic features would provide the most pronounced generation. 
The hypothesised topographic scatter of storm-generated near-inertial waves have, to our knowledge, not been studied before and thus it is unknown whether they constitute a significant contribution to the ocean internal wave field and deep ocean mixing.
To investigate this phenomenon, we run a numerical ocean model with idealised wind forcing and bathymetry.
Under various topographic and wind forcing scenarios, we calculate the energy flux associated with these waves, and compare the magnitude of energy flux to that of surface-generated near inertial waves excited directly by the wind.

The next section reviews the theory of \cite{gill_behavior_1984}, and demonstrates the excitation of a bottom flow from the dephasing mechanism.
In section~\ref{section: method}, we describe the model setup, as well as the method we used for separating the waves of interest from the rest of the flow. 
Section~\ref{sec:results} includes a qualitative demonstration of these bottom scattered waves, followed by a quantitative analysis of the wave energy for various perturbation experiments. 
Finally, we conclude in section~\ref{section: discussion} by arguing that this mechanism could be responsible for some of the unexplained observations of excess bottom mixing and vertical shear.

\section{Theory}

\cite{gill_behavior_1984} modelled the ocean's immediate response to the passage of a storm as an initial velocity $u_0$ in the mixed layer, with zero velocity deeper in the water column. 
This initial condition can be thought of as an infinite set of orthonormal vertical modes each with a different vertical wavenumber $m_n$, which destructively interfere below the mixed layer.
As these modes necessarily propagate at different speeds, they rapidly de-phase from one another, resulting in finite flows over the entire ocean depth.

To demonstrate this mechanism, consider an idealised hydrostatic 2D model of the ocean with a mixed layer of depth $h_m$, total depth $H$, and constant stratification $N$, just after the passage of a storm with characteristic horizontal length scale $L$ and associated wavenumber $k = 2 \pi / L$. 
The initial surface response to the storm when $t = 0$ is given by an initial mixed-layer velocity with a maximum of $u_0$, while the interior is at rest; that is,
\begin{equation}
\label{eq:u0}
u(x,z,t=0) = u_0 \cos(kx) \mathcal{H}(z + h_m),
\end{equation}
\noindent where $z=0$ corresponds to the ocean's surface and $\mathcal{H}$ is the Heaviside function, equal to 0 when $z<-h_m$. Here, we follow the typical method of decomposition of the linearised Boussinesq equations into a set of vertical modes with boundary conditions $\mathrm{d}u / \mathrm{d} z = 0$ at $z = 0$ and $z = -H$ \citep{olbers_ocean_2012}.
For an arbitrary stratification profile $N(z)$, the eigenvalue problem cannot typically be solved analytically, but for the case of constant stratification~$N$ the eigenfunctions take the form $\phi_n(z) = \sqrt{2/H}\cos(n \pi z /H)$. The eigenvalues $m_n$ correspond to the vertical wavenumbers $m_n=n \pi/H$ of the solution and are related to the horizontal wavenumber $k$ (set by the spatial structure of the wind forcing) and frequency $\omega_n$ via the dispersion relation for internal waves 
\begin{equation}
  \omega_n^2 = f^2 + N^2 \frac{k^2}{m_n^2}.
  \label{eq:dispersion}
\end{equation}

The baroclinic velocity can be expanded in vertical modes as:
\begin{align}
    u(x, z, t) = \sum^{\infty}_{n = 1} a_n  \phi_n(z) \cos(k x - \omega_n t),
\end{align}
where the amplitudes $a_n$ are obtained by projecting the initial velocity~\eqref{eq:u0} onto the $n$th mode:
\begin{align}
    \label{eq:a_n}
    a_n & = \int^0_{-H} \phi_n(z) \, u_0 \mathcal{H}(z+h_m) \, \mathrm{d}z .
\end{align}
The barotropic part of the solution is the vertical mean of the initial condition. Note that this example represents the surface forcing not as a single isolated storm, but as an infinite wind band that repeats spatially at the characteristic length scale $L$. 
For the purposes of this idealised model, the horizontal structure is unimportant, so we simply consider the location $x = 0$ to investigate the interior response. 

Each of the vertical modes propagates at a different frequency and phase speed according to the dispersion relation for internal waves~\eqref{eq:dispersion}
meaning that lower modes propagate faster in the horizontal.
Initially, the linear combination of modes with amplitudes prescribed by~\eqref{eq:a_n} sum to give $u_0$ in the mixed layer and zero at depth. But as time progresses, the different propagation speeds mean that the vertical modes de-phase from one another, and flow develops at depth.

\begin{figure}[h]
    \centering    
    \includegraphics[width=\columnwidth]{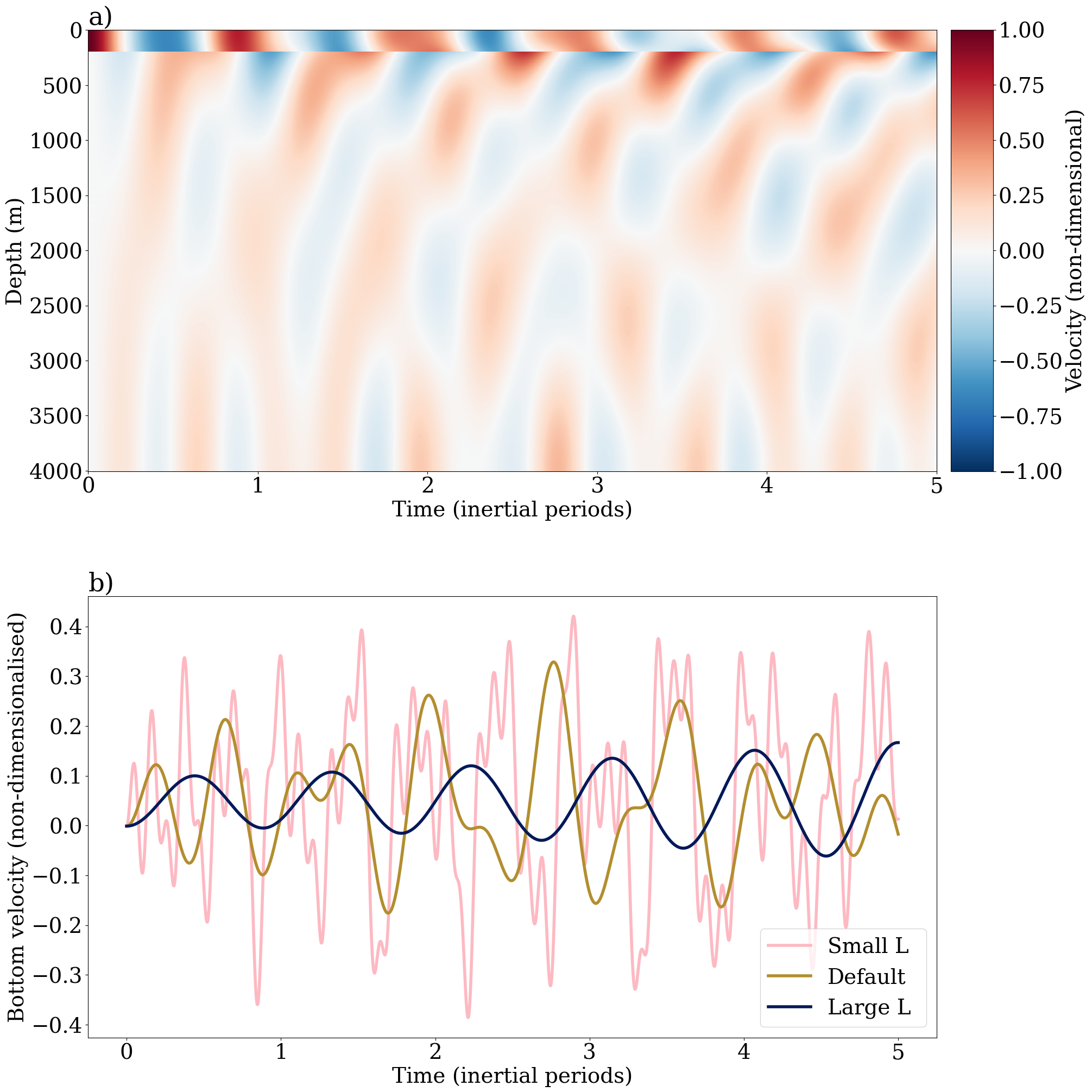}
    \caption{a) Evolution of the initial surface layer velocity $u(x,z,t=0)$ for a reference case of $N = 20|f|$ and a horizontal scale of $L=2\pi/k=100\;\mathrm{km}$. b) The bottom velocity response corresponding to the reference conditions, as well as for storms with decreased ($20~\mathrm{km}$) and increased ($1000~\mathrm{km}$) horizontal scale $L$. Velocities have been non-dimensionalized against $u_0$, which in this case is $0.1\;\mathrm{m}\;\mathrm{s}^{-1}$ over a $200\;\mathrm{m}$ thick mixed layer $h_m$.}
    \label{fig:theory}
\end{figure}

The manner in which the dephasing happens depends on the differences in frequency $\omega_n$ and in turn on stratification $N$ and length scale $L$.
Figure~\ref{fig:theory} shows the vertical structure of the solution at $x=0$ for $h_m = 200\;\mathrm{m}$, $H = 4000\;\mathrm{m}$, $|f| = 10^{-4}\;\mathrm{s}^{-1}$ and $N = 20|f|$ over 5 inertial periods. 
In this case, flow develops throughout the water column, including at the bottom, within 10\% of an inertial period.
Figure~\ref{fig:theory}b shows that varying $L$ affects both the amplitude and frequencies of resultant bottom flow, with the solution tending towards purely inertial oscillations as $L$ increases. 
As propagating waves can be generated at topography with super-inertial oscillations, of particular interest are the stronger super-inertial motions that can be seen in the cases with smaller horizontal length scale.

The thickness of the surface layer $h_m$ plays a role in re-distributing the energy between vertical modes, but does not affect their fundamental dephasing behaviour. This simple model demonstrates that a wind-induced surface perturbation in velocity --- of the correct scale --- can almost instantaneously generate super-inertial oscillatory flow at the sea floor.
Existing theory for internal tides can now be applied in evaluating the likely impact of the wind-induced oscillatory flow on secondary wave generation --- or scatter --- at the seafloor.
The conversion rate $C$ of energy from an oscillatory flow of amplitude $U_0$ to buoyancy frequency $N$, topography height $h$, and horizontal wavenumber $k$, based on linear wave theory, is given by \citep{jayne_parameterizing_2001}:
\begin{equation}
    C = \frac{1}{4} \rho_0 U_0^2 h^2 N k ,
    \label{eq:jayne}
\end{equation}
where $\rho_0$ is the reference density. 
Although this mechanism differs from internal tides in that the oscillatory flow excited by winds is not of constant amplitude or frequency, \eqref{eq:jayne} holds for our study to leading order if $U_0$ refers instead to the characteristic amplitude of the wind-induced velocity. 
Our numerical results are evaluated against this scaling relation in section~\ref{section: discussion}.

\section{Numerical model and analysis methodology}
\label{section: method}
\begin{figure*}[h]
  \centering  
  \includegraphics[width=0.7\textwidth,angle=0]{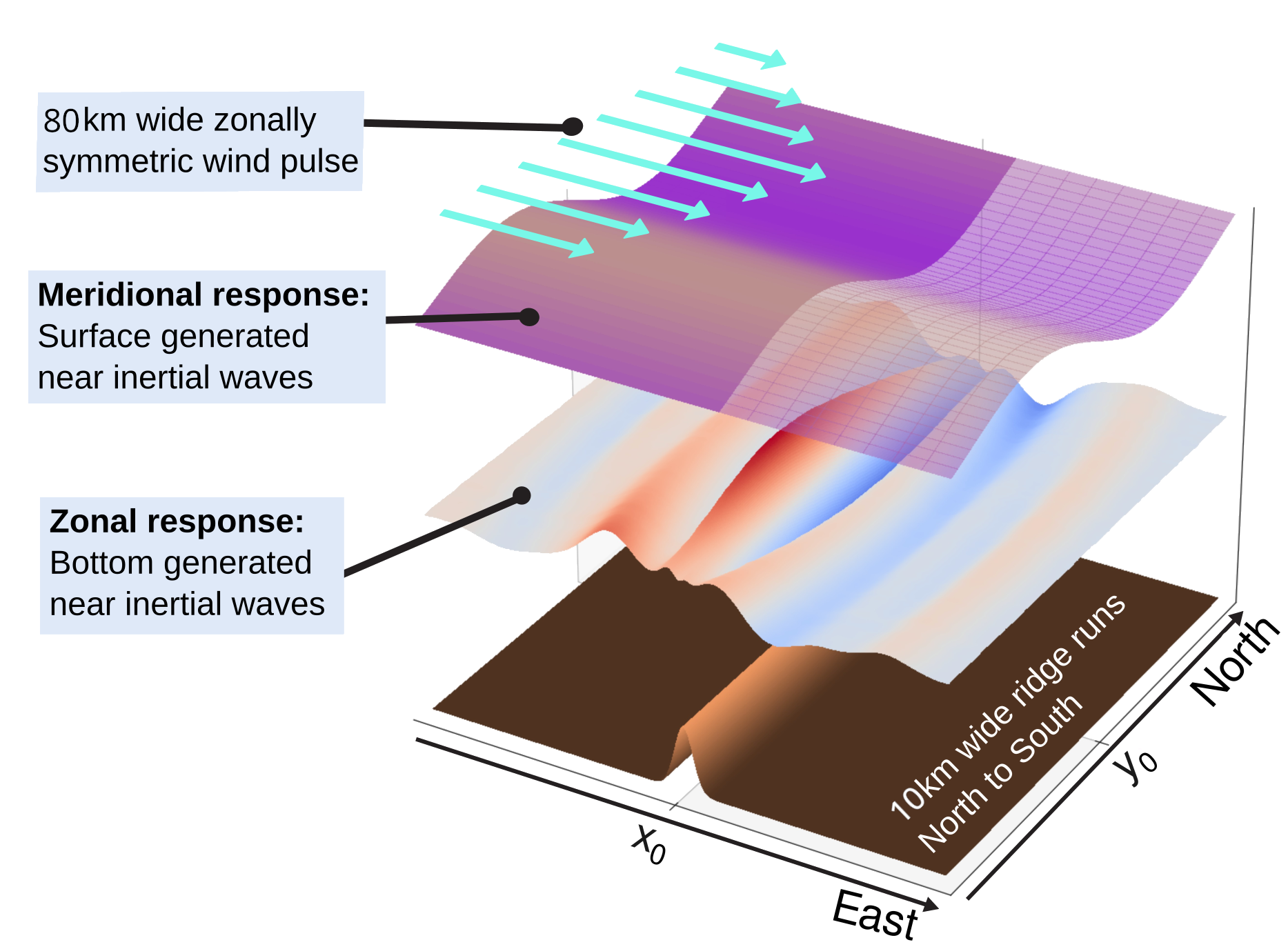}
  \caption{A schematic of the model configuration, including the isopycnal displacements associated with both the surface and bottom generated near-inertial wave responses. Note that the vertical scales of the waves have been exaggerated, and that two layers are shown separately and in different colours for illustrative purposes, having been filtered using the zonal symmetry of the surface near-inertial wave response. The zonal and meridional centre lines as referred to in the text are denoted by $x_0$ and $y_0$ respectively.}
  \label{fig:schematic}
\end{figure*}

To study topographic near-inertial wave generation mechanism, we design an idealised numerical experiment as shown in Fig.~\ref{fig:schematic}. 
The Modular Ocean Model~6 (MOM6; \citet{adcroft_gfdl_2019}) is run in an isopycnal configuration with 21 layers at 2~km horizontal resolution, total depth of 4000~m and horizontal extent of 4000~km by 4000~km. 
The zonal and meridional boundaries are re-entrant, but the domain is large enough that internal waves do not return to the region of study around the ridge during the simulations. 

The  top isopycnal layer mimics a mixed layer and has a thickness of $50~\mathrm{m}$ in the default case; the second layer below mimics a thermocline and initially has a thickness of $150~\mathrm{m}$; the remaining 19~layers below have an initial thickness of $200~\mathrm{m}$. A Coriolis parameter of $f = - 10^{-4}$~s$^{-1}$ corresponding to the Southern mid-latitudes is selected. The top layer has a density of $1018\;\mathrm{kg}/\mathrm{m}^3$, followed by a large jump in density with an equivalent buoyancy frequency $N = 200|f|$ to simulate a thermocline. 
The remaining 19 layers have a linearly increasing density to give a constant buoyancy frequency of $40|f|$, an approximation of a typical buoyancy frequency observed beneath the thermocline \citep{reagan_james_r_world_2021}.

A challenge in studying wind-generated topographic near-inertial waves is that, even in a highly idealised domain, the waves exist in the presence of other flows in the inertial band.
These include non-propagating inertial oscillations as well as near-inertial waves generated directly from the wind forcing that have not undergone any interactions with topography.
To simplify the separation of wind-induced topographic near-inertial waves from these other near-inertial and inertial flows, the experiment was designed with a high degree of symmetry. 
The wind pulse is zonally symmetric, so that any zonally propagating waves are a result of interactions with the topography, and so this zonal component contains the waves of interest. 

For our control experiment, a zonally symmetric 1000~m tall Gaussian ridge with full width at half maximum (FWHM) $W = 10\;\mathrm{km}$, running north to south through the domain, is placed in the domain centre. 
These parameters for the cross-ridge and vertical scales are chosen to be consistent with a typical seamount  \citep{wessel_global_2010}.

The model is initialised from rest and forced by a zonally-uniform 10-hour pulse of zonal wind stress given by
\begin{equation}
\tau(y,t) = 
    \begin{cases}
        ~~\,\tau_0 \, e^{-y^2 / 2 \sigma^2}\sin^2\left(2\pi t/T_\tau\right) & 0 \leq t \leq  T_\tau / 2, \\
        -\tau_0 \, e^{-y^2 / 2 \sigma^2} \sin^2\left( 2\pi t/T_\tau\right) & T_\tau / 2 < t \leq T_\tau,   \\
        ~~\,0 & T_{\tau} < t,
    \end{cases}
    \label{eq:windprofile}
\end{equation}
where $\tau_0$ is the peak wind stress, $L_\tau = 2\sqrt{2\log{(2)}} \sigma$ is the FWHM of the wind stress, and $T_\tau$ the forcing duration, with default values of $T_\tau=10\;\mathrm{hours}$ and $L_\tau = 80\;\mathrm{km}$, consistent with observed tropical cyclone size distributions \citep{zhang_era5_2023}. The default wind stress value $\tau_0 = 10\;\mathrm{N}\,\mathrm{m}^{-2}$ corresponds to the strength of a category~4 tropical cyclone with surface wind speed of~$63\;\mathrm{m}\,\mathrm{s}^{-1}$.
The wind pulse is focused at the meridional-centre of the domain, and the eastward and westward oscillation serves to force the surface layer without inducing a significant time-mean flow, providing a robust idealised model of the process of interest (i.e., high-frequency wind forcing by storms).
A small westward time-mean flow, less than $10^{-6}\;\mathrm{m}\,\mathrm{s}^{-1}$, is still present, but this is too small to meet the $|f| < k U_0$ condition required to support topographic lee waves.

In choosing to filter the waves by symmetry, a drawback is that our wind event lacks zonal gradients, restricting the divergence of the wind field to the meridional gradient. 
This formulation means that, while near-inertial waves are expected to be generated due to this sharp, unidirectional gradient, the magnitude of these waves may be significantly reduced compared to those generated by the winds of a true cyclone, which has both strong zonal and meridional gradients. 
However, as the scattered wave energy scales linearly with the incident waves to first order (see~\eqref{eq:jayne}), the comparison between the incident and scattered wave-fields remains as an important metric, even if the energy fluxes themselves are under-represented.

Employing the experiment's zonal and meridional symmetries to separate incident and scattered waves, we calculate the wave energy fluxes propagating outwards through a square centred on $x_0$, $y_0$ with side lengths $D$.
The surface-generated NIW energy flux, $E_{S}$, corresponds to the energy flux propagating meridionally through the northern and southern sides, whereas the energy flux of the NIWs resulting from interactions with topography, $E_T$, correspond to the wave energy fluxes through the eastern and western boundaries of the square. 
Energy fluxes are the product of the velocities $u$ or $v$ normal to the grid-cell face, and pressure $P$.
The symmetry of the domain ensures that the energy of the topographically-generated near-inertial waves that are the focus of this study is captured within $E_T$.

To isolate the velocities $u$ and pressures $P$ required to calculate $E_T$, we subtract reference values from $2000\;\mathrm{km}$ west of the ridge, such that $u' = u - u_{\rm far}$ and likewise for $P$.
Multiplying the perturbation velocity and pressure, and summing over the opposing sides of the square yields the zonal $E_{i,T}$ and meridional $E_{i,S}$ energy fluxes at the $i$th layer:
\begin{align}
\label{eq:eflux}
E_{i,T}(y,t) & = \left[P_i' u_i'\right]_{x = \frac{D}{2}} \; - \; \left[P_i'u_i'\right]_{x = -\frac{D}{2}}, \\
E_{i,S}(x,t) & =\left[P_i v_i\right]_{y = \frac{D}{2}} \; - \; \left[P_i v_i\right]_{y = -\frac{D}{2}}.
\end{align} 
These energy fluxes are then integrated in space and time to determine the total energy leaving the square meridionally and zonally. 
The quantities of interest are the vertical energy flux from the storm directly into the surface-generated NIWs, and the vertical energy flux emanating upwards from the topography.
Rather than directly calculating these surface and topographic vertical energy fluxes, which is challenging in an isopycnal model with layers intersecting with the topography, we instead assume that these energy fluxes are equivalent to the corresponding horizontal energy fluxes.
This assumption means that the calculated energy fluxes will be underestimated by any dissipation or breaking that occurs during the horizontal propagation, but in this idealised model with no background flow and otherwise smooth topography, dissipative mechanisms are minimal. 
To allow time for the signal to propagate to the edges of the square, the analysis period extends $90$ hours beyond the end of the storm but is still normalised by the $10$ hour storm duration.

With this assumption, we then define our  average vertical energy flux quantities $\overline{E_T}$  and~$\overline{E_S}$:
\begin{align}
    \label{eq:eflux_mean}
    \overline{E_{T}} &= \frac{1}{T_{\tau} D W}\sum^N_{i = 1} h_i \int^{\frac{D}{2}}_{-\frac{D}{2}} \int^T_0 E_{i,T} \,\mathrm{d}t \mathrm{d}y, \\
    \overline{E_{S}} &= \frac{1}{T_{\tau} D L_\tau}\sum^N_{i = 1} h_i \int^{\frac{D}{2}}_{-\frac{D}{2}} \int^T_0 E_{i,S} \,\mathrm{d}t \mathrm{d}x, 
\end{align}
where $h_i$ is the $i$th layer thickness, $T$ is the total simulation duration, and $T_\tau$ the duration of the storm.
The experiment is run for $90$ hours after the end of the storm, giving a total duration of $T = T_\tau + 90\;\mathrm{hrs}$, equivalent to $100$ hours in the default case.
Rather than divide by depth $H$ to get the average energy flux through the sides of the square, factor~$1/W$ in~\eqref{eq:eflux_mean} converts the depth-integrated (summed over each isopycnal layer) horizontal energy flux to an equivalent vertical flux over the topography, providing a quantity in units of power per unit area of the topography (usually expressed in $\mathrm{W} / \mathrm{m^2}$).
Likewise, the factor $1/L_{\tau}$ converts the energy flux in meridional waves into a downward flux per unit area of the storm. 
These quantities are referred to as the surface and topographic energy fluxes throughout the study. 

The choice of side length $D$ for our square of energy flux integration needed to be close enough to the ridge that waves could travel a distance $D / 2$ well before any waves would reflect back from the sides of the domain.
Conversely, $D$ must be larger than all of the ridge widths tested.
Through testing several values of $D$, we found that $500$ km was large enough to avoid any large amplitude, non-linear effects near the ridge, yet remaining within reasonable computational constraints concerning the required domain size. 

\section{Results}
\label{sec:results}

\begin{figure*}[h]
  \centering
  \includegraphics[width=0.81\textwidth, angle=0]{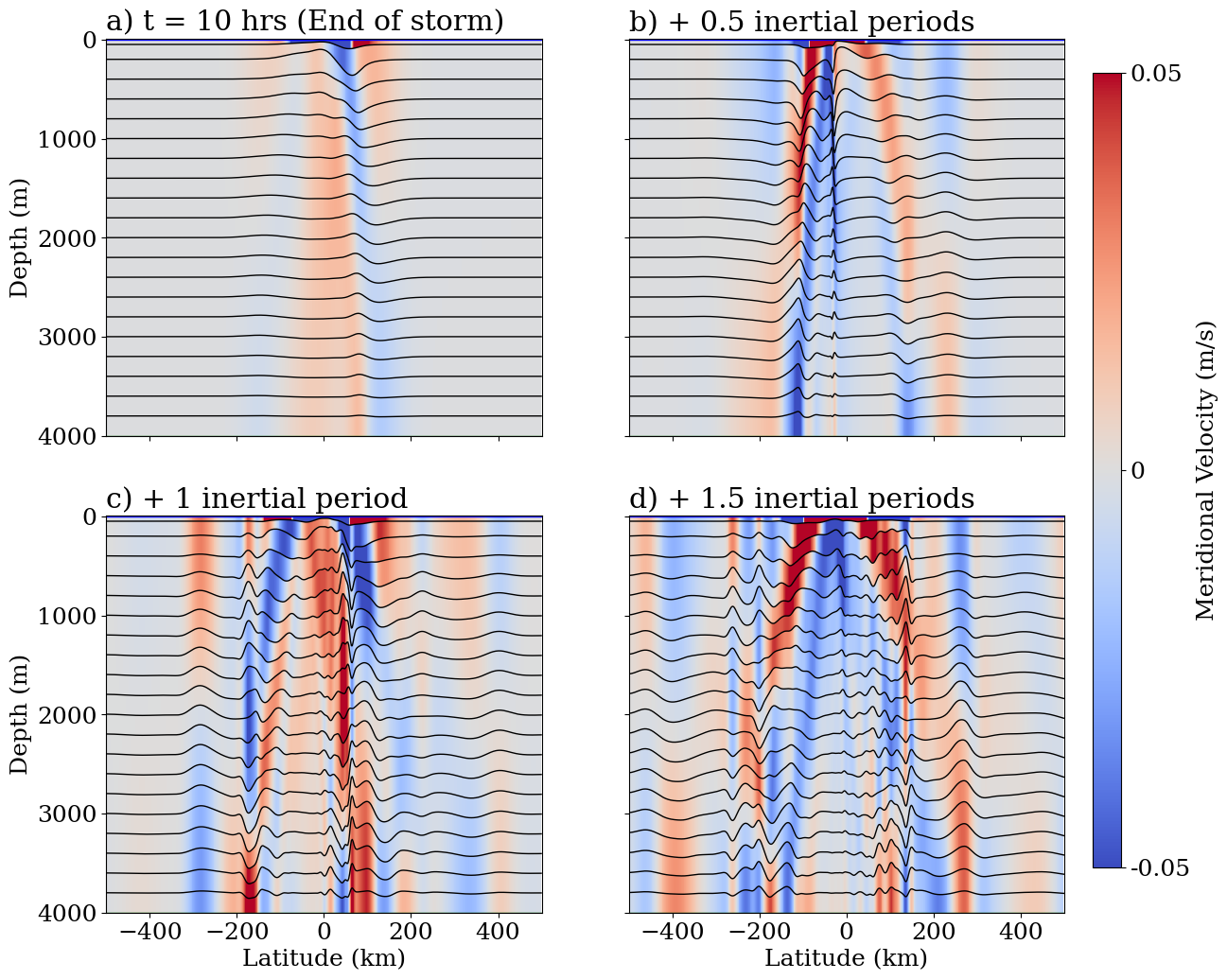}
  \caption{Meridional transect showing meridional velocity and exaggerated isopycnal displacements associated with wind-generated NIWs. Isopycnal displacements are not to scale; they are exaggerated for visualization purposes.}
  \label{fig:stopmotion_merid}
\end{figure*}
We first provide a qualitative demonstration of the proposed rapid scatter of near-inertial waves at topography.
Fig.~\ref{fig:stopmotion_merid} shows the meridional velocities and isopycnal height anomalies associated with the meridionally propagating near-inertial waves generated directly at the surface.
Internal waves are immediately visible through the full water column, similar to the analytic theory shown in Fig.~\ref{fig:theory}.
As time progresses, we see that the mode~1 waves (characterised by the single zero crossing) dominant in panel~(a) give way to third and fourth vertical modes in panels~(c) and~(d). 
This behaviour is consistent with the modal dephasing theory in that the faster, lower modes appear first, which propagate more rapidly from the centre of the storm.
By inspecting the phase shifts of individual vertical modes and the beam-like structures, it is clear that the frequency is near-inertial. 
Importantly to the generation of waves from topography, Fig.~\ref{fig:stopmotion_merid} shows significant motions at the seafloor immediately after the storm event.

\begin{figure*}[h]
  \centering
  \includegraphics[width=0.81\textwidth]{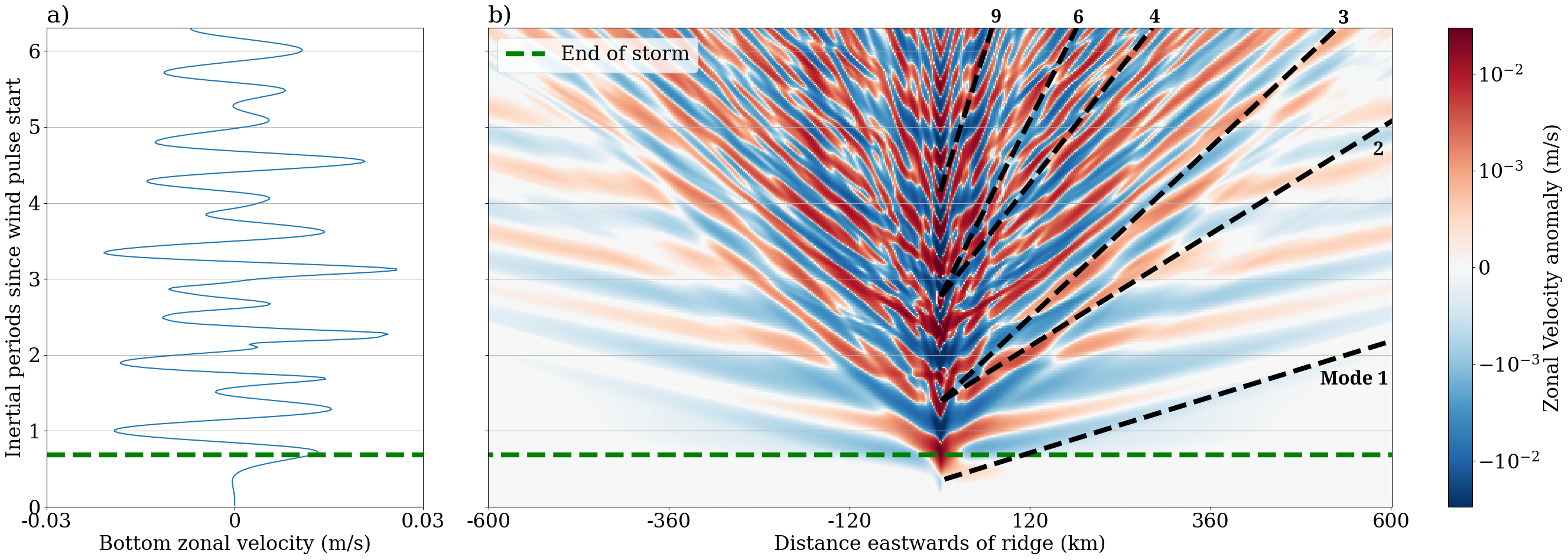}
   \caption{Panel a)  Bottom zonal velocity 1000 km West of ridge, b) Hovmöller diagram of zonal velocity anomalies 5 layers from the bottom in the vicinity of the ridge with the approximate group speeds for vertical modes 1, 2, 3, 4, 6, and 9. Both panels show velocities 200 km north of the storm centre, as in the bottom row of Fig.~\ref{fig:stopmotion}.}
  \label{fig:hovmoller}
\end{figure*}

These oscillatory motions are more clearly visible in Fig.~\ref{fig:hovmoller}a, which shows bottom velocities directly beneath the wind forcing similar to the 2D analytic theory in Fig.~\ref{fig:theory}b.
Here, the model shows an immediate bottom response with a clear super-inertial signal in the bottom flow, which enables the generation of propagating internal waves in the presence of appropriate topography. 
A Hovmöller diagram in Fig.~\ref{fig:hovmoller}b shows a signal in the resulting zonal velocity propagating away from the ridge, including several superimposed vertical modes characterised by their different group speeds, fanning out from the topography.

\begin{figure*}[h]
  \centering
  \includegraphics[width=0.81\textwidth, angle=0]{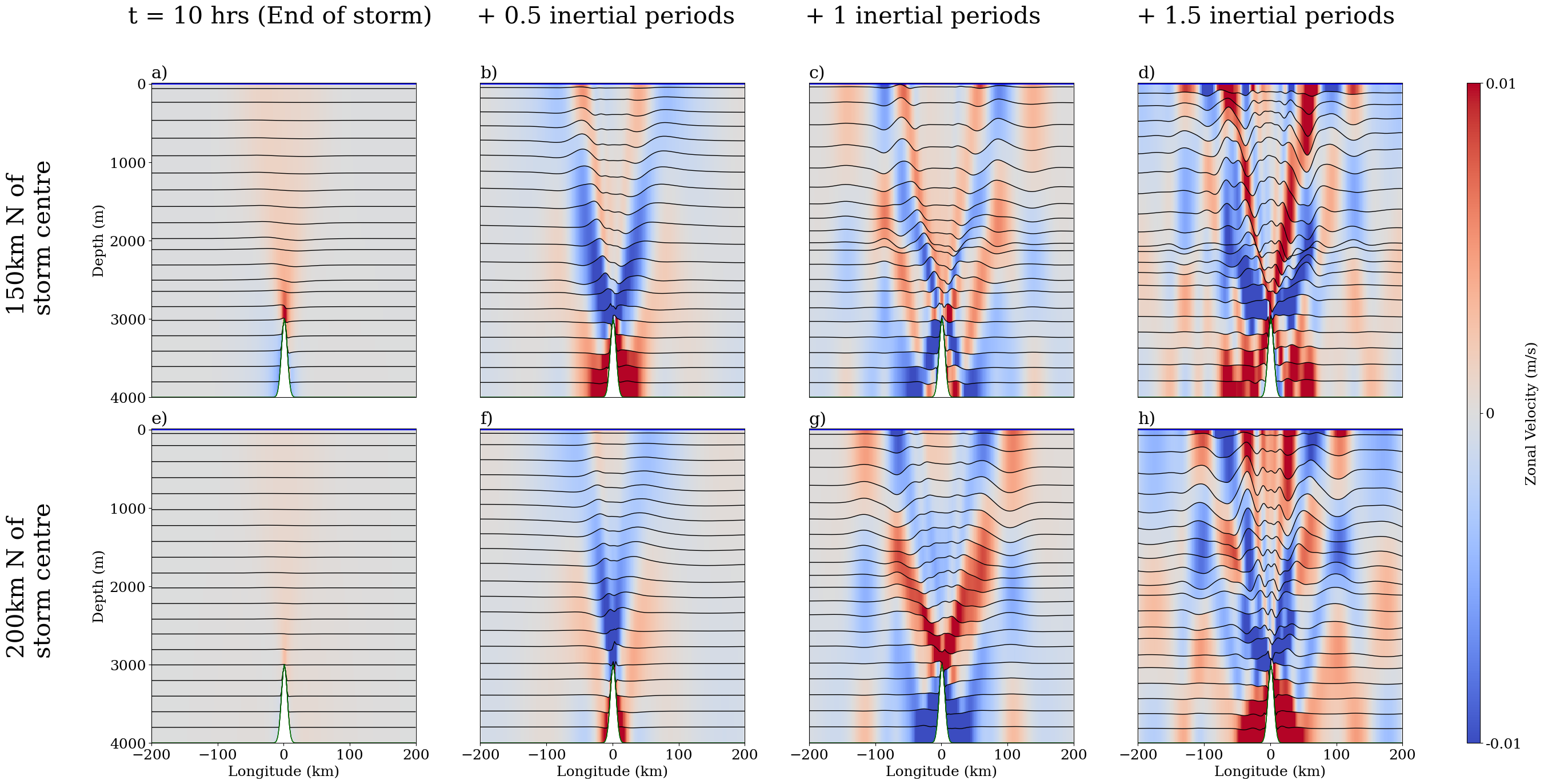}
   \caption{Zonal transect showing zonal velocity (color) and exaggerated isopycnal displacements (lines) associated with wind-generated topographic NIWs from our reference numerical simulation at four different time snapshots and two locations north of the storm. }
  \label{fig:stopmotion}
\end{figure*}

To visualise the bottom scattered waves, the velocity and interfacial height anomalies along two zonal transects are plotted in Fig.~\ref{fig:stopmotion}.
Here, as in \eqref{eq:eflux}, the anomalies are calculated with respect to their far-field values to remove the effects of non-topographically-generated waves. 
In all panels, clear beams can be seen propagating upwards and outwards from the top of the ridge.
The two rows, corresponding to locations $150$ and $200\;\mathrm{km}$ northwards from the centre of the wind forcing, show meridional variations to the waves structure and phase.
This is due to the time taken for the dephasing signal to propagate outwards from the storm, as well as the meridional structure of the storm itself.

\begin{figure}
  \centering
  \includegraphics[width=\columnwidth]{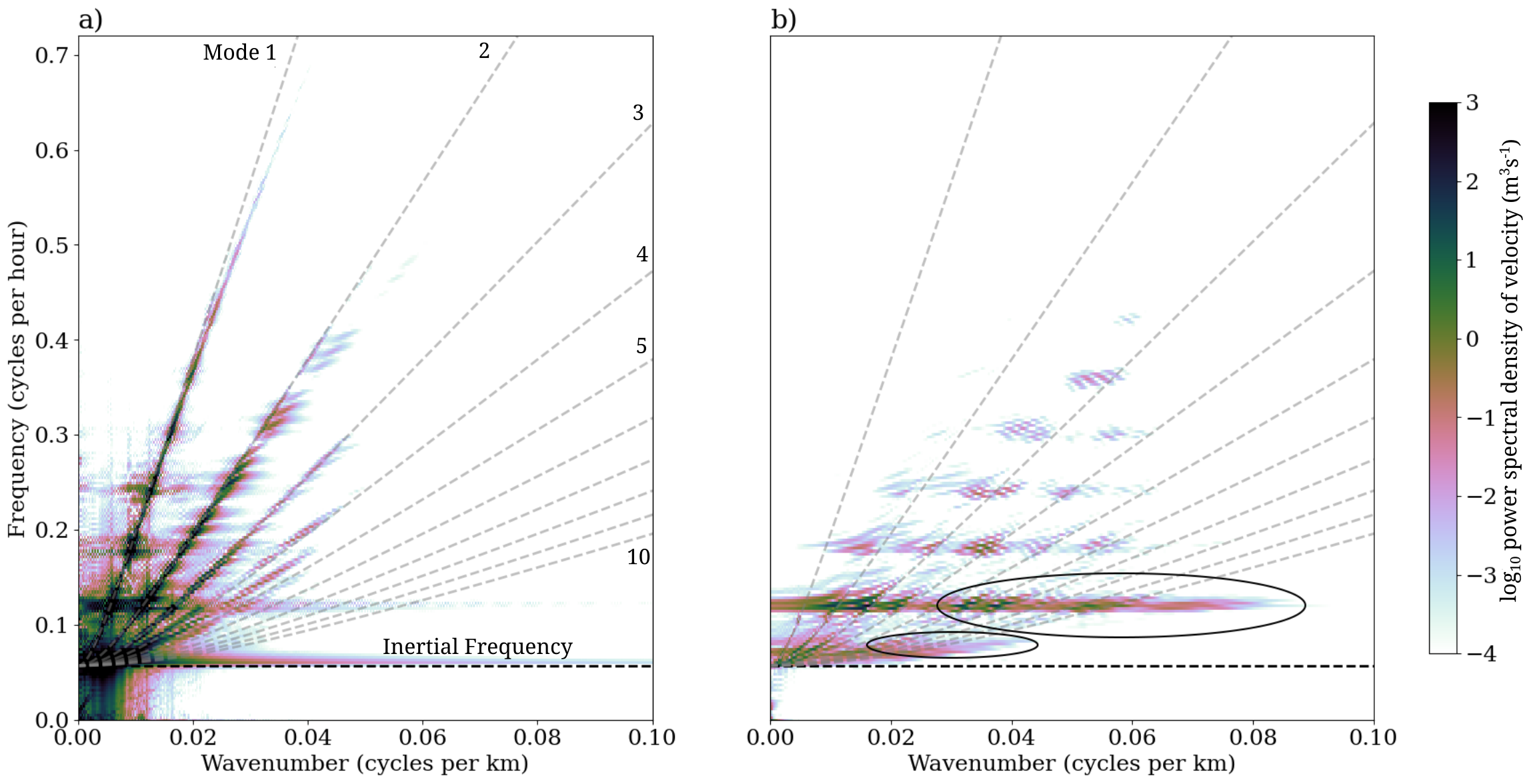}
  \caption{Depth averaged power spectra with the first 10 analytic vertical near-inertial modes overlaid in dashed grey of: a) the meridional velocity associated with the incident surface-generated near inertial waves b) the zonal velocity associated with the topographically scattered waves. Both spectra are taken over the first 10 inertial periods of the experiment, and at the same spatial positions as the corresponding waves shown in figures \ref{fig:stopmotion_merid} and \ref{fig:stopmotion}. The ovals in panel b highlights additional energy in modes 6-10 not present in panel a.}
  \label{fig:spectrum}
\end{figure}

To assess the role of these bottom scattered waves in redistributing energy to higher vertical modes and smaller scales, we take the power spectra of the velocities associated with the meridional and zonal waves in Fig.~\ref{fig:spectrum}.
The prominent vertical modes of the surface-generated, meridional waves visible in Fig.~\ref{fig:spectrum}a are consistent with the dispersion relation~\eqref{eq:dispersion}, and show a dominant first vertical mode, with energy also spread through modes 2-5. 
The spectrum of the scattered waves, shown in Fig.~\ref{fig:spectrum}b, shows bands of power close to harmonics of $|f|$, with local maxima where these bands intersect vertical modes. 
Compared to the incident meridionally propagating waves, the scattered, zonally propagating waves have a very weak mode~1, and the circled regions in panel~b show significant energy present in modes~6-10. 
Thus energy is distributed further towards higher modes in the scattered zonal waves than the incident meridionally propagating waves. 
This difference in the modal energy distribution shows that the bottom scatter of NIWs generated at the surface by storms can play a role in downscaling the energy to higher vertical modes, and in turn smaller horizontal scales.

Figure~\ref{fig:spectrum}a shows that the incident meridional waves have most energy close to $|f|$, but with significant energy in the first harmonic.
This enhanced first harmonic can be explained by the wind time scale of the storm -- at $10$ hours, the dominant frequency is close to $2|f|$, a choice to ensure that most of the wind energy spectrum is super-inertial, in order to generate the most wave activity. 
In contrast to Fig.~\ref{fig:spectrum}a, in Fig.~\ref{fig:spectrum}b, we see that the first harmonic is more energetic than the near-inertial band, suggesting that this dephasing mechanism for bottom scatter is more efficient for $2|f|$.

This observation is consistent with the strength of the superinertial component of the bottom oscillations compared to the purely inertial in both Fig.~\ref{fig:hovmoller}a and Fig.~\ref{fig:theory}. 
Here, higher harmonics of $|f|$ are identifiable as the additional, consistently spaced zero crossings between inertial periods.
These higher harmonics in Fig.~\ref{fig:hovmoller}a are modulated by by a near-inertial signal of comparable amplitude, much like the two smaller wind forcing scale cases in Fig.~\ref{fig:theory}. 
The excitation of waves at $2|f|$ is likely due to higher harmonics dephasing more readily than lower: since the spatial scales imposed by the topography remain constant between the $|f|$ and $2|f|$ waves, their group speeds remain the same, resulting in the faster dephasing of the $2|f|$ harmonic due to it having more cycles per unit of distance.  
This explains the prominence of the $2|f|$ harmonic in this case of a transient storm system generating a rapid bottom response.

The waves shown in Figs.~\ref{fig:stopmotion} and~\ref{fig:hovmoller} have characteristics consistent with the proposed mechanism, and Fig.~\ref{fig:spectrum} demonstrates their potential for the downscaling of wave energy. 
Thus, these results serve as a qualitative demonstration of wind-induced topographically generated near-inertial waves. 
Having qualitatively demonstrated the rapid scatter of storm-generated NIWs at topography, we now quantify the wave energy, in order to assess in what circumstances, if any, these waves might pose a significant source of bottom intensified wave activity.

\subsection{Wave Energy Flux}
\label{section:energy_flux}
Having established the mechanism for the rapid topographic scattering of storm generated near-inertial waves in our configuration, we now proceed to calculate the corresponding energy fluxes of the incident and scattered waves.
From~\eqref{eq:eflux}, the time averaged energy flux in our reference configuration (Figs.~\ref{fig:stopmotion_merid}, \ref{fig:stopmotion}, and~\ref{fig:hovmoller} above) associated with the wind-generated topographic NIWs is $0.13\;\mathrm{mW}\,\mathrm{m}^{-2}$, compared to $1.9\;\mathrm{mW}\,\mathrm{m^{-2}}$ for the wind-generated surface NIWs.
As discussed in section \ref{section: method}, these energy flux values are an underestimate of what might be expected in the real ocean due to the simplified, zonally symmetric wind forcing that enables our clean wave separation.
Thus, a more important metric is in comparing the relative sizes of the incident and scattered wave energy fluxes.
With the control set of parameters -- namely with a ridge $1000\mathrm{m}$ high and $10\mathrm{km}$ wide, and $63~\mathrm{ms}^{-1}$ winds lasting for $10$ hours -- the zonally propagating, scattered waves have a time-averaged energy flux 7\% the size of that of meridionally propagating surface-generated NIWs.

\subsection{Perturbation experiments}
\label{section:pert_expts}
We run experiments varying both the properties of the storm and ridge to assess which parameters the energy flux is most sensitive to and determine the conditions under which the topographically-generated waves are most significant. 

\begin{figure*}[h]
\centering
\includegraphics[width = \textwidth]{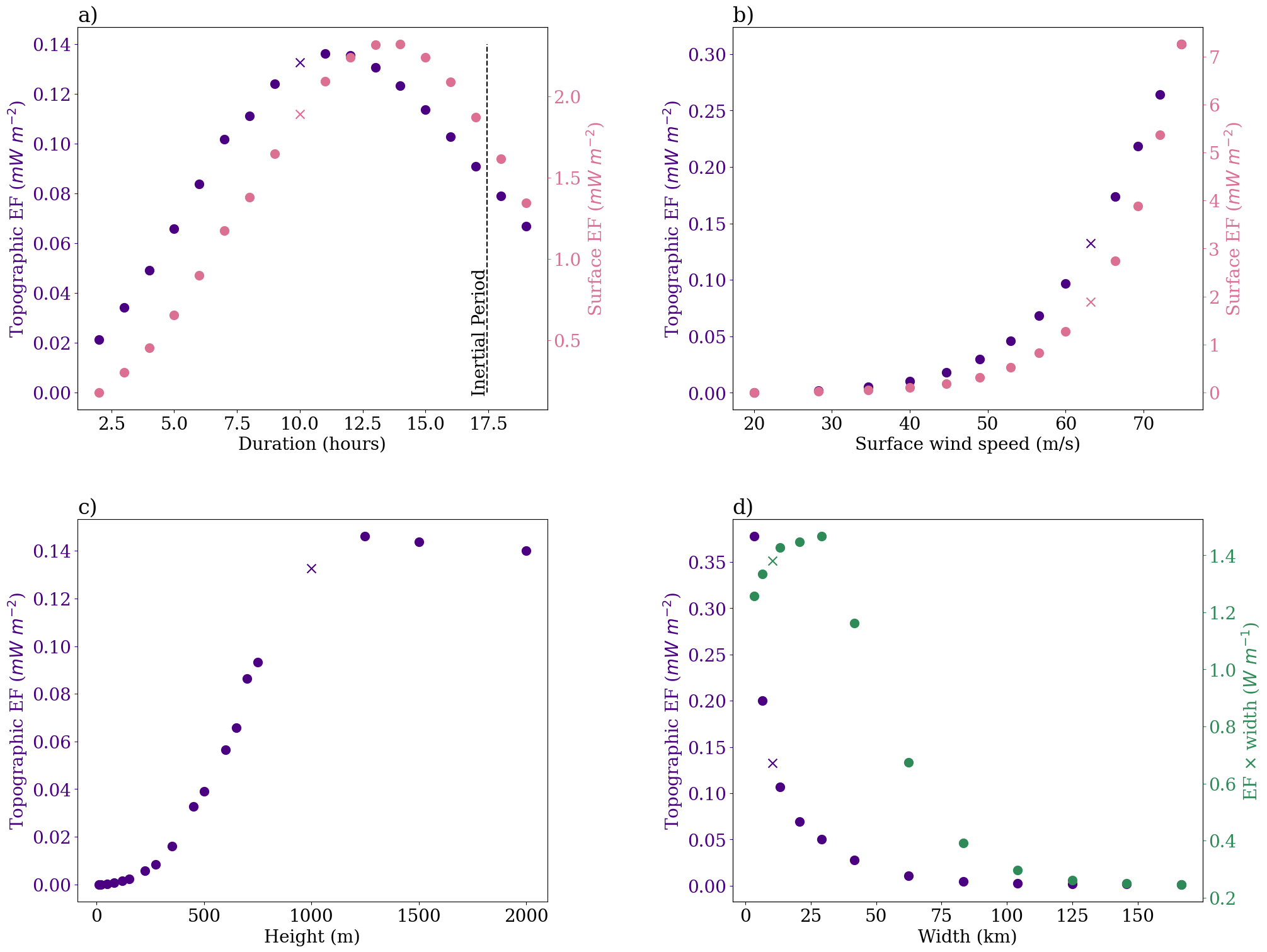}
\caption{The energy flux of the wind-driven topographic near-inertial (zonal) waves when varying a) forcing duration; b) surface forcing wind speed; c) ridge height; d) ridge width. The default values are shown with an $\times$. The energy flux for the wind-driven surface-generated near-inertial (meridional) waves is shown in panels a) and b) only, as it is unaffected by varying the shape of the topography.}
\label{fig:perturbations}
\end{figure*}

When altering the duration of the wind forcing pulse (Fig.~\ref{fig:perturbations}a) there is a peak in both topographic and surface energy fluxes when the forcing duration is less than the inertial period, i.e., $2\pi / |f| \approx 17.5$~hours.
This result is consistent with expectations for surface-driven near-inertial waves, as internal wave generation is only supported by super-inertial forcing frequencies.
Forcing durations right at the inertial frequency would overlap significantly with the sub-inertial band, whereas a more rapid surface forcing would contain less energy at the highly resonant inertial period. 
The offset peaks in Fig.~\ref{fig:perturbations}a) are likely due to the exciting of the $2|f|$ harmonic in the scattered waves, shown in Fig.~\ref{fig:spectrum}b.
This is due to the significant $2|f|$ component of the bottom velocity (see Fig.~\ref{fig:hovmoller}a) which in turn is caused by the faster dephasing of the higher harmonic waves than the near-inertial. 
 
The strength of the wind forcing pulse was altered in Fig.~\ref{fig:perturbations}b, ranging from winds expected during a mild storm up to a category~5 tropical cyclone. 
Here, both the topographic and surface wave energy fluxes increase quadratically (with correlation coefficient $0.95$).
This quadratic relationship matches the wind work applied at the surface, which scales quadratically with the surface wind speed.

The second series of tests conducted relate to the shape of the topography (Fig.~\ref{fig:perturbations}c,d). 
As expected from equation~\eqref{eq:jayne}, the energy flux scales quadratically (with correlation coefficient $0.95$) with ridge height up to $750~\mathrm{m}$, or just shy of a quarter of the water column depth.
The energy flux then appears to saturate, likely due to the larger topography blocking the cross-ridge flow, and thereby damping wave generation \citep{winters_topographic_2014}. 

Finally, Fig.~\ref{fig:perturbations}d shows the sensitivity of topographic energy flux when varying the ridge width. 
Here, the scaled energy flux, as in~\eqref{eq:eflux}, is shown alongside the product of the energy flux and width. 
This plot is designed to show the way that the total wave energy changes, as well as energy flux, which scales inversely to ridge width.
Here, we see that the energy flux per unit area (Fig.~\ref{fig:perturbations}d, left axis) is greatest for the narrowest ridge, but that the total wave energy per metre meridionally (Fig.~\ref{fig:perturbations}d, right axis) peaks for a $30\;\mathrm{km}$ wide ridge.
This effect is due to the maximum topographic slope becoming shallower than the angle of the wave beam, suppressing the downward propagating part of the wave-field visible in Fig.~\ref{fig:stopmotion} (result not shown). 
In~\eqref{eq:jayne}, $k$ refers to the dominant wavenumber of the ridge, which in our case is $k \sim 1/W$.
This is consistent with the our results in Fig.~\ref{fig:perturbations}, whereby the energy flux per unit area is greater for steeper and narrower topography.

\section{Discussion and conclusions}
\label{section: discussion}

We demonstrate a novel mechanism for the generation of wind-driven near-inertial waves at topography within hours of the passage of a storm. 
These waves are generated by the rapid dephasing of horizontally propagating full-depth wave modes induced by a transient wind event, and interaction of the resulting bottom flow with topography to radiate waves. 
This generation mechanism is showcased using an isopycnal model forced by a wind pulse that mimics a storm. 
The idealised model setup is carefully constructed to enable a clean separation of the scattered waves from surface-generated near-inertial waves and to eliminate any significant time-mean flow that could generate lee waves.

The perturbation experiments shown in Fig.~\ref{fig:perturbations} reveal that the variation of wave energy fluxes in our experiment is consistent with the existing internal wave theory. 
With our control experiment, namely a $1000\;\mathrm{m}$-high, $10\;\mathrm{km}$ wide ridge in a $4000\;\mathrm{m}$ deep ocean, with an $80\;\mathrm{km}$ wide (FWHM) category 4 tropical cyclone overhead, the topographically scattered near-inertial waves have energy fluxes of $0.13\;\mathrm{mW}/\mathrm{m}^2$, that is, $7\%$ the energy flux in surface-generated NIWs in the same experiment. 
Given that our simplified storm under-represents surface NIW generation, we focus on the comparison between the surface-generated and topographically scattered waves, and consider these energy flux estimates as a lower bound. 

With the most narrow topography that we test, the maximum energy flux attained is $0.39\;\mathrm{mW}/\mathrm{m}^2$, or $18\%$ of the surface-generated NIWs. 
However, as such a prominent ridge or seamount would be uncommon, we consider as a more typical range, the energy flux ratios from the strength and duration perturbation experiments (Fig.~\ref{fig:perturbations}a,b). 
From these two experiments, the ratios of the topographically generated to the directly surface-generated NIW energy fluxes range from 5\% at the longest duration and strongest storm, to 10\% at the shortest and weakest storms.
This range is consistent with the 6\% of surface NIW energy estimated to reach the seafloor in the work by  \cite{jouanno_dissipation_2016}, except that with our proposed mechanism this energy is communicated rapidly (within 100 hours of the storm) to the bottom.

Given that the mechanism described in this study relies on specific and short-lived circumstances, namely a storm event near prominent topography, it is not expected to be a major contributor to the internal wave field in a globally averaged sense. 
Instead, the mechanism could be important for explaining intermittent locally enhanced internal wave generation at the seafloor and deep-ocean mixing.
Recent studies have identified evidence for unexplained enhancement of internal wave activity at the seafloor.
\cite{van_haren_challenger_2020} finds anomalously large bottom near-inertial oscillations, mixing and deep overturning at $3000\;\mathrm{m}$ depth $1000\;\mathrm{km}$ away from a category~4 tropical cyclone. 
This bottom signal's first detection and intensification happened simultaneously with the approach of the cyclone, pointing to a rapid (on the order of hours or a few days) communication of the storm's energy to the deep ocean.
Similarly, in another observational study, \cite{morozov_inertial_2008} show a rapid bottom response coinciding with the arrival of two cyclones, in addition to the downward propagating wave-packet which takes two weeks to reach the seafloor (see Figs.~11 and~12 by \cite{morozov_inertial_2008}).
A modelling study of storm-induced flows shows enhanced near-inertial energies at the seafloor within 5 days of the passage of a storm, several days ahead of the downward propagating wave packets (see Fig.~9 by \cite{jouanno_dissipation_2016}).
\cite{jouanno_dissipation_2016} note that while this bottom NIW generation is linked to the storm, the associated energies are two orders of magnitude smaller than the surface-generated NIWs and were thus not investigated further.
However, our perturbation experiments suggest this ratio of surface to bottom generated NIW energy flux could be an order of magnitude larger than was found by \cite{jouanno_dissipation_2016}, warranting further investigation.
We argue that the mechanism proposed in the study could at least partly account for these findings in the existing literature, but further study is needed to clarify the extent to which this may be the case.

Here, we studied the generation of wind-driven topographic near-inertial waves in a highly idealised domain to cleanly separate them from other flow features.
An important next step is to identify the waves in more realistic simulations. 
However, delineating topographic near-inertial waves from other flow factors is difficult due to their generation by the dephasing of the vertical modes of the surface response, which itself includes NIWs of similar frequencies and spatial scales.
Our experiments were carefully designed to separate the wind-driven topographic near-inertial waves from other flow features as cleanly as possible.
Consequently, this caused the underestimation of the near-inertial wave energy imparted by the storm and limited the scope of our study to simple dynamics, topographic features, and surface forcing profiles.
To investigate topographic near-inertial waves in a more realistic numerical simulation or with observations, these intricacies of disentangling them from other flow features will need to be handled carefully. 
One option could be to exploit the difference in vertical propagation between the topographic and wind-generated near-inertial waves. 
If a filter were used to separate the upwards and downwards propagating signals, perhaps following the spectral approach employed by \cite{olbers_closure_2017} or \cite{waterhouse_global_2022}, one could identify areas of increased vertical energy flux beneath storm systems. 
However, one would still need to remove the background internal wave fields, among other oscillatory signals that could obscure the waves of interest. 
To obtain a stronger signal, a statistical approach could be employed, e.g., averaging over many storm events in several locations above regions of varying topographic prominence.
At each location, measurements of the bottom mixing could be taken both in the absence and presence of an overhead storm. 
Future studies could employ this approach on high-resolution model output or observations. 

As shown in Fig.~\ref{fig:spectrum}, the scatter of storm-generated NIWs at topography can redistribute energy towards higher vertical modes, and smaller horizontal scales. The higher vertical modes visible in Fig.~\ref{fig:spectrum}b suggests that these scattered waves may be especially important for abyssal mixing due to their role in energy downscaling. However, our study was designed to maximise both wave generation and propagation so that the scattered wave-field was easier to isolate and study and therefore mechanisms for wave breaking and dissipation were intentionally either removed or suppressed. Hence we made no attempt to estimate abyssal mixing generated by this mechanism. 

In addition to the spatial downscaling, the scattered waves tend to higher frequencies, particularly the prominent first harmonic of $|f|$. The reason for this is that the power spectrum of the bottom velocity from the dephasing of the incident waves in Fig.~\ref{fig:hovmoller} contains a $20\%$ stronger $2|f|$ peak than $|f|$ (figure not shown), likely because the higher harmonics dephase more quickly and so are more prominent in the post-storm bottom velocity. As higher frequency internal waves exhibit less vertical shear, they are less prone to breaking, so this could somewhat offset any increase to mixing attributed to the spatial downscaling. Higher harmonics in the downscaling of near-inertial waves by eddies was observed in \cite{danioux_emergence_2011}. In this study, authors found that the $2|f|$ harmonic dominated $|f|$ at depth, stating that the mechanism explaining this shift in frequency in depth remains unresolved. If the role of eddies in downscaling these near-inertial waves also involves the dephasing of vertical modes, then the results of our study could help to explain these findings.

The results of our idealised study suggest that the scatter of storm-generated NIWs at topography may contribute to energy downscaling and in turn deep ocean mixing, but the exact extent and location of enhanced bottom mixing due to our proposed mechanism remains an open question. 
It remains possible that in more realistic conditions the waves would immediately break, become trapped, or otherwise dissipate close to the generation site, which was intentionally avoided in our study to better measure the energy fluxes. 
Further study is therefore needed to understand how these waves would behave in the real ocean.
If wind-forced topographic near-inertial waves are responsible for energy downscaling and intensified near-bottom mixing, then one could assess their impact indirectly by measuring the wavenumber spectra and bottom mixing rates at other intersection points of topographic features and storm systems. 
This could be done either with existing moorings, or by analysing model outputs of high resolution ocean models with more realistic forcing and bathymetry.
The advantage of using high-resolution ocean models is that the effects of tides could easily be removed, so enhanced internal wave generation or breaking at the bottom in the presence of storms could more easily be teased out.
It would also be worthwhile to apply these analysis techniques to the data of the previously mentioned studies \citep{van_haren_challenger_2020,morozov_inertial_2008,jouanno_dissipation_2016} which found evidence for bottom NIW activity immediately after a storm, to investigate whether this mechanism is indeed consistent with these observations.

In conclusion, we have studied a novel internal-wave generation mechanism whereby the rapid dephasing of a wind-driven internal waves modes results in bottom flow over topographic features. 
The energy flux associated with these wind-driven topographic near-inertial waves reached~$0.2\;\mathrm{mW}\,\mathrm{m}^{-2}$, a value $10\%$ the size of surface-generated NIWs in the same domain. 
By separating the topographic near-inertial waves from other dynamical features in our idealised experiments, we found that wave energy scales in a way consistent with existing internal wave theory. 
Further study of topographic near-inertial waves in ocean observations and realistic models would help us clarify how these waves might fit into the broader internal wave spectrum, and our picture of abyssal mixing. 

\acknowledgments
Computational resources were provided by the National Computational Infrastructure at the Australian National University, which is supported by the Commonwealth Government of Australia.
A.J.B.~acknowledges their PhD stipend provided by the Australian Research Training Program. 
We acknowledge funding from the Australian Research Council under DECRA Fellowship DE210100749 (N.C.C.), the Centre of Excellence for Climate Extremes CE170100023 (A.J.B., A.M.H., and C.J.S.), and the Centre of Excellence for the Weather of the 21st Century CE230100012 (A.M.H. and N.C.C.).
We would like to thank the two anonymous reviewers and the editor for their constructive comments that greatly improved the manuscript.

\datastatement

{
The model configuration setup, post processing, analysis and figure rendering code is publicly available at \url{github.com/ashjbarnes/topographic-NIWs}, and replicated in a Zenodo repository at \url{https://doi.org/10.5281/zenodo.15493445}. 
The MOM6 source code can be found at \url{github.com/mom-ocean/MOM6}.
}

\end{document}